\documentclass[twocolumn,showpacs,preprintnumbers,amsmath,amssymb,prl,superscriptaddress]{revtex4-1}
\usepackage{graphicx}
\usepackage{dcolumn}

\usepackage{booktabs}
\usepackage{bm}
\newcolumntype{.}[1]{D{.}{.}{#1}}
\usepackage{float}
\usepackage{color}
\usepackage{hyperref}
\hypersetup{
	breaklinks=true,
	pdfnewwindow=true,      
	colorlinks=true,       
	linkcolor=blue,          
	citecolor=blue,        
	filecolor=blue,      
	urlcolor=blue,          
}
\begin{document}
\title{The $\epsilon$-$\zeta$ Transition in Solid Oxygen}
\author{S. F. Elatresh}
\email[Electronic address:\\]{sabri.elatresh@kfupm.edu.sa, bonev@llnl.gov}
\affiliation{Physics Department and Interdisciplinary Research Center for Intelligent Secure Systems, King Fahd University of Petroleum and Minerals, Dhahran, Saudi Arabia}
\author{V. Askarpour}
\affiliation{Department of Physics, Dalhousie University, Halifax, NS B3H 3J5, Canada}
\author{S. A. Bonev}
\email[Electronic address:\\]{sabri.elatresh@kfupm.edu.sa, bonev@llnl.gov}
\affiliation{Lawrence Livermore National Laboratory, Livermore, CA 94550, USA}


\date{\today}

\begin{abstract} 
 The structure of solid oxygen has been studied at pressures from 50 to 140~GPa using static structure search methods and molecular dynamics simulations with density functional theory and a hybrid exchange functional. Several crystalline structures with space group symmetries  {\it Pnma}, {\it P}\,2$_{1}${\it /m}, {\it Pm} and {\it P}\,6$_3$/{\it mmc} have been identified as candidates for the $\zeta$ phase of oxygen at 0~K. Within the hybrid exchange functional framework and at 300~K temperature, {\it Pm} is shown to be energetically most favorable above 111~GPa. A comparison with experimental X-ray diffraction, spectroscopic and superconductivity measurements is provided for all competing structures. 

\end{abstract}

\maketitle

Oxygen is in many ways a unique element: fundamental, abundantly available on Earth, and the only known diatomic molecule that carries 
a magnetic moment ~\cite{Lundegaard:2006jl}. Its phase diagram is a good example of the abundance of interesting physical properties, which emerge under compression.  Multiple phases have been discovered  at low pressure, including the $\gamma, \alpha, \beta, \delta, \eta,$ and  $\varepsilon$ phases~\cite{FREIMAN20181,PhysRevLett.74.4690,Freiman20041,Lundegaard:2006jl,doi:10.1107/S0365110X62002248,PhysRevB.37.5364,doi:10.1063/1.3118970,PhysRevB.29.1387,Schiferl:a22092,PhysRevB.65.172106,PhysRevLett.93.055502,doi:10.1021/j100366a020,Shimizu:1998,Zhu17012012,PhysRevLett.112.247201,PhysRevLett.91.265503}, which are molecular 
 and the structural differences among them are in the relative arrangements of the O$_2$ molecules.  At room temperature, solid oxygen 
 exists in the non-magnetic~\cite{PhysRevLett.94.218701} insulating $\varepsilon$ phase between 10~GPa and 96~GPa~ \cite{PhysRevLett.97.085503,Lundegaard:2006jl}, above which it undergoes
 an insulator-to-metal transformation to the $\zeta$ phase~\cite{PhysRevLett.74.4690}. The latter transition appears to be gradual and completed by 110~GPa~\cite{PhysRevLett.88.035504}.

The $\varepsilon$-phase, which remains stable across a wide range of pressures, has attracted a lot of interest,  encompassing both theoretical investigations~\cite{Meng19082008,PhysRevLett.80.5160,PhysRevB.61.6145,PhysRevLett.88.205503,Sabrithesis,2013APS..SHK.B3005E,Anh2019,doi:10.1021/acs.jctc.5b00017,PhysRevB.92.085148,OCHOACALLE201582,Ramirez-Solis_PCCP_2017} and experimental examinations employing X-ray diffraction~\cite{Johnson:mo0090,PhysRevLett.74.4690,PhysRevLett.88.035504,Fukui21385} and 
spectroscopic measurements~\cite{PhysRevB.54.R15602,PhysRevLett.83.4093,PhysRevB.61.8801,:/content/aip/journal/jcp/86/10/10.1063/1.452547,Akahama_JJAP_2019}. 
Powder~\cite{PhysRevLett.97.085503,PhysRevLett.74.4690} and single-crystal~\cite{PhysRevLett.88.035504} X-ray diffraction 
studies indicated that the $\varepsilon$ phase has a monoclinic {\it C}\,2{\it /m} symmetry with a primitive cell 
with a single O$_8$ cluster consisting of four molecules.  A theoretical GW analysis of the $\varepsilon$ phase reported that the structural transformation to $\varepsilon$  is accompanied by an insulator-metal transition at 51.7~GPa~\cite{PhysRevB.77.092104} while a later GW study suggested a transition pressure of about 100~GPa~\cite{PhysRevB.78.132101}.
 
Although the structure of the $\varepsilon$ phase is experimentally known ~\cite{Lundegaard:2006jl,PhysRevLett.97.085503,Meng19082008}, that of the $\zeta$ phase remains a subject of continued debate. Raman measurements~\cite{PhysRevB.68.100102} demonstrated that the $\zeta$ phase is molecular with at least four $O_2$ pairs in the primitive unit cell, while conductivity data~ \cite{Shimizu:1998} suggested superconductivity below 0.6~K. Although powder X-ray diffraction~\cite{PhysRevLett.74.4690}, and single-crystal X-ray and Raman studies~\cite{PhysRevLett.102.255503} proposed the structure of $\zeta$-O$_2$ to be monoclinic with symmetry {\it C}\,2{\it /m} (hereafter referred to as $\zeta$-{\it C}\,2{\it /m} to distinguish it from $\varepsilon$-{\it C}\,2{\it /m}, which we will
refer to as $\varepsilon$-O$_8$), Raman and single-crystal X-ray diffraction experiments~\cite{PhysRevLett.88.035504,PhysRevB.68.100102} seemed to imply that the $\varepsilon$ and $\zeta$ phases are not isostructural.
{\it Ab initio} crystal structure prediction methodology~\cite{PhysRevB.76.064101} performed within the generalized gradient approximation (GGA) to density dunctional theory (DFT) also proposed the structure of $\zeta$-O$_2$ to be {\it C}\,2{\it /m}. However, several discrepancies exist between the theoretical predictions and experimental data.

We recall that the lowest-enthalpy structure for the $\varepsilon$ phase was initially predicted to be {\it Cmcm} and not the experimentally determined O$_8$  ($\varepsilon$-{\it C}\,2{\it /m}). In addition, DFT calculations within GGA initially suggested that the $\varepsilon$-$\zeta$ transition occurs at approximately 40 GPa, in contrast to the experimental value of 96 GPa.
These discrepancies were recently resolved by independent theoretical studies~\cite{C9CP05267D, Ochoa-Calle2015, PhysRevB.92.085148, Ram-Sols2018} where the researchers emphasized the significance of employing accurate exchange-correlation (XC) functionals beyond the GGA.

The present study is focused on the $\epsilon$-$\zeta$ transition and determining the structure of the $\zeta$ phase. Several new candidate structures for the $\zeta$ phase are found using the Universal Structure Predictor Evolutionary Xtallography (USPEX) code~\cite{10.1063/1.2210932,GLASS2006713,OGANOV200695,LYAKHOV20101623,Oganov2011,LYAKHOV20131172}.
In light of the the recent studies~\cite{C9CP05267D, Ochoa-Calle2015, PhysRevB.92.085148, Ram-Sols2018} elucidating the importance of using  accurate exchange-correlation functionals, we have performed  hybrid exchange calculations within the Heyd-Scuseria-Ernzerhof approximation
(HSE06) ~\cite{doi:10.1063/1.2204597}. Based on them, a new, HSE06-modified phase diagram is proposed.  Furthermore, molecular dynamics simulations of the candidate structures are performed and their HSE06-corrected Gibbs free energies computed for a final relative stability ordering of the candidate structures. 
 
Electronic structure calculations and structural relaxations were performed within DFT using the Vienna {\it ab initio} simulation package (VASP)~\cite{PhysRevB.47.558,PhysRevB.54.11169}.  The generalized gradient approximation (GGA) along with projector augmented Wave (PAW) Perdew-Burke-Ernzerhof (PBE) pseudopotential in which the 2{\it s}$^2$2{\it p}$^4$ states were treated as valence states with a core radius of 1.1~a.u. was employed~\cite{PhysRevB.59.1758,prl96Perdew}. The geometry was optimized using the conjugate gradient algorithm. The Brillouin zone (BZ) ${\bf k}$-point sampling was performed with the Monkhorst–Pack scheme~\cite{PhysRevB.13.5188}. The electronic density-of-states (DOS) was calculated on a 20$\times$20$\times$20 grid in the BZ by the tetrahedron method with Bl\"{o}chl corrections~\cite{PhysRevB.49.16223}. Raman modes were calculated by density functional perturbation theory as implemented in the ABINIT package~\cite{Gonze20092582}. 

\begin{figure} [t!]
 \centering 
\includegraphics[width=0.5\textwidth, clip]{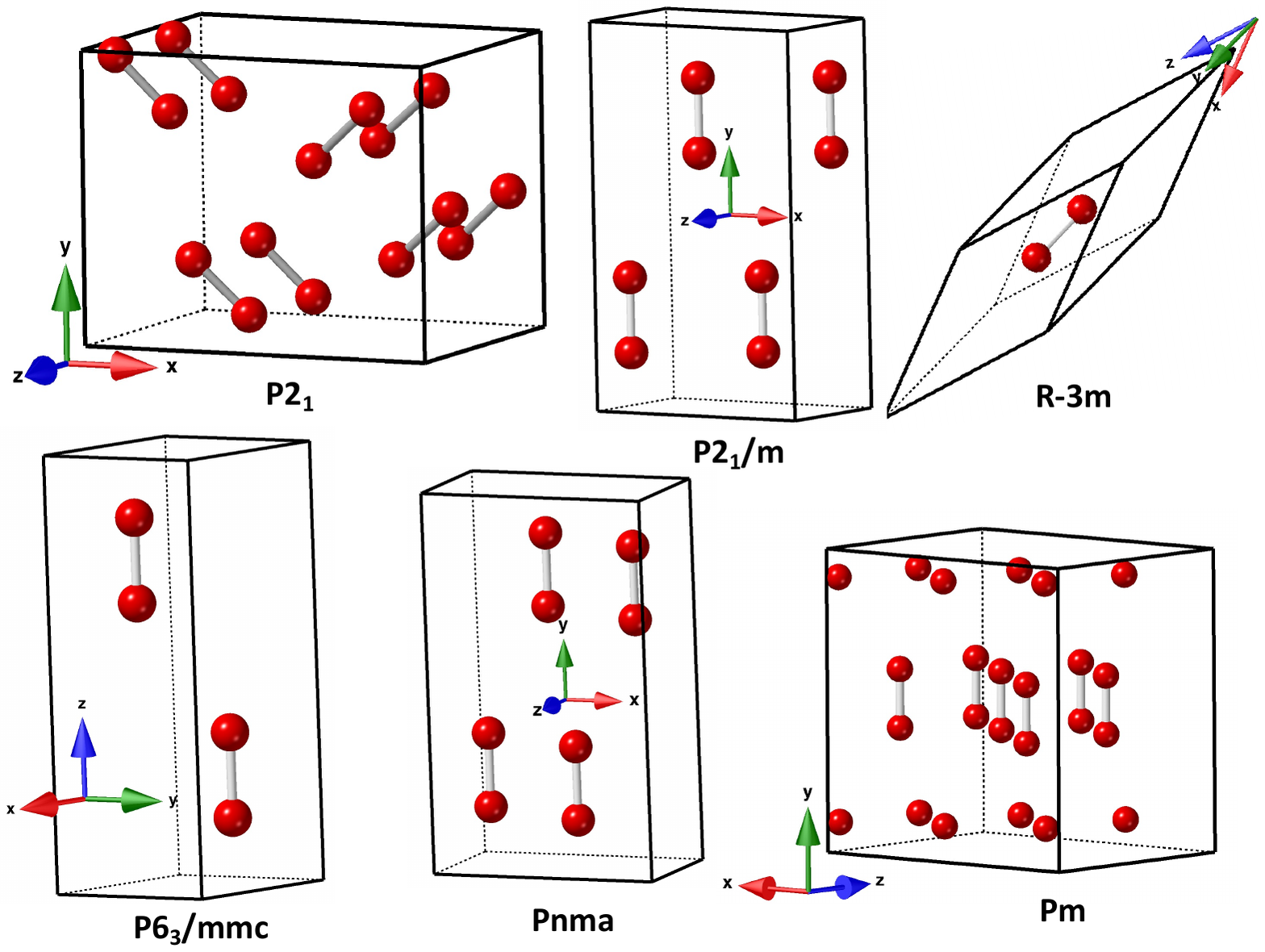}
\caption{Candidate crystal structures for solid oxygen at 90 GPa based on static, 0~K, DFT-PBE searches. Shown are the newly found 
 {\it P2}${_1}$, {\it P}2${_1}${\it /m}, {\it R}-3{\it m}, {\it P}6$_3$/{\it mmc}, {\it Pnma}, and {\it Pm} structures.\label{All-St}}
\end{figure}

\begin{figure}[t]
\includegraphics[scale=0.31,trim = 1mm 1mm 1mm 2mm, clip]{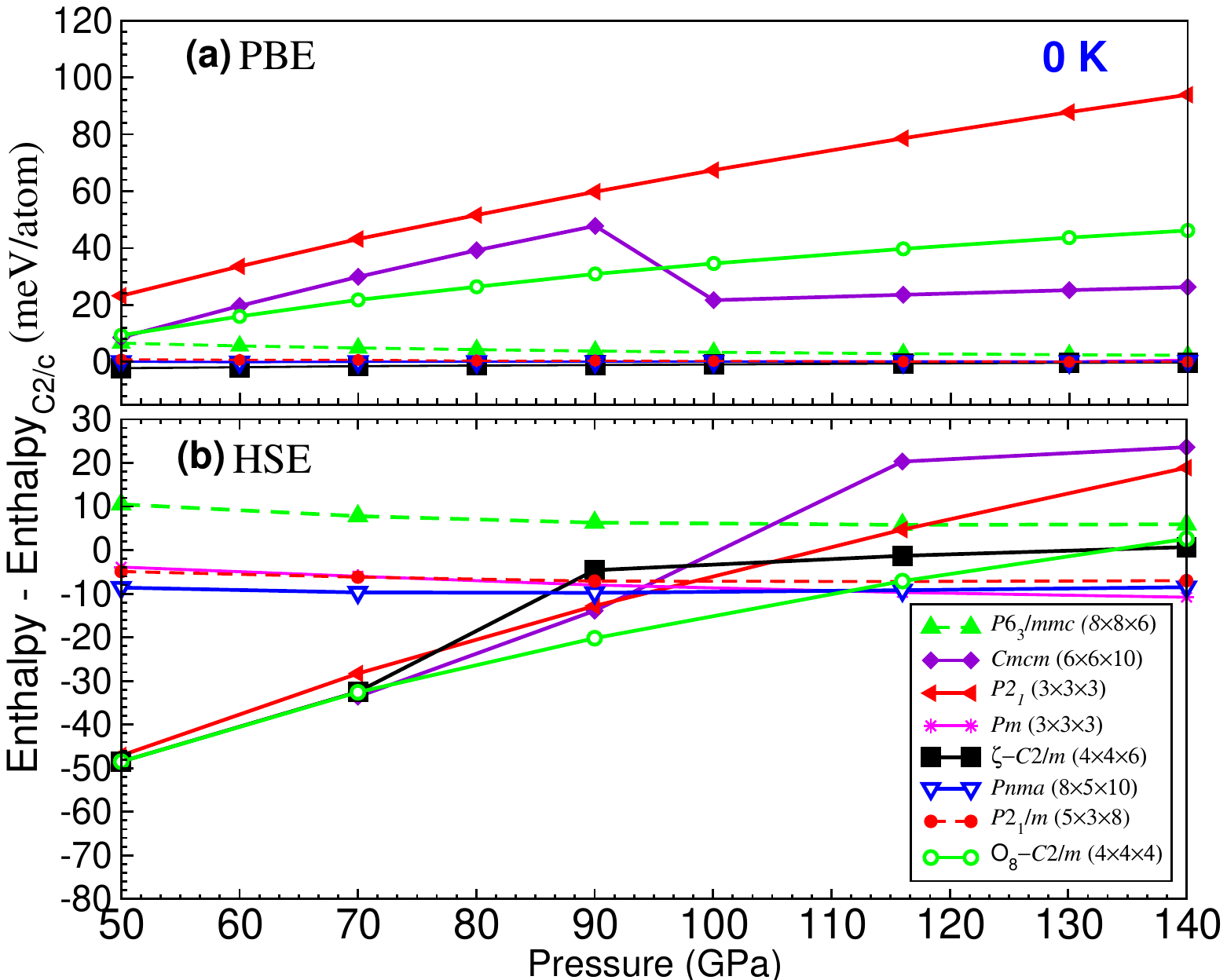}
\caption{Enthalpies of various structures of oxygen relative to the {\it C}\,2{\it /c} phase relaxed within (a) PBE (16$\times$16$\times$16 ${\bf k}$-point mesh) and (b) HSE06 (8$\times$8$\times$4).\label{H-P_0K}}
\end{figure}

Search for new structures were conducted for 16 and 24 atoms at 35~GPa and 150~GPa using the USPEX code~\cite{10.1063/1.2210932,GLASS2006713,OGANOV200695,LYAKHOV20101623,Oganov2011,LYAKHOV20131172}. In each case, starting from an initial population of 50 random structures, variable-cell relaxation was performed by VASP using a cutoff energy of 800~eV, a 4$\times$4$\times$4 Monkhorst-Pack grid and Methfessel-Paxton smearing~\cite{PhysRevB.40.3616} with a width of 0.1~eV. In addition to several of the lowest-enthalpy structures that automatically moved on to the next generation, 70\% of each generation was used to create the next generation by hereditary and mutation. After over 30 generations, the lowest-enthalpy structures (Fig.~\ref{All-St}) were further relaxed  with an energy cutoff of 1000~eV, denser BZ sampling grids (listed for each structure in the caption of Fig.~\ref{H-P_0K}(a)), and force convergence requirement of 1~meV/$\AA$. Based on convergence tests, these parameters were sufficient to converge the total PBE energies to better than 1~meV/atom. 

It is well know that GGA (DFT) underestimates the electronic band gap. Therefore, the HSE06 functional~\cite{doi:10.1063/1.2204597} was employed to describe the molecular and electronic properties of solid oxygen. The electron correlation interaction and the long-range part of the exchange interaction were maintained at the PBE level while 25\% of Hartree-Fock exact exchange and 75\% of the PBE exchange were combined to form the short-range part of the electron-electron exchange interaction. To  carry out the demanding HSE06 calculations within a reasonable time, the number of k-points were lowered (see  Fig.~\ref{H-P_0K}(b) caption), which introduced an estimated error of 2~meV/atom in the calculated energy values.

The space groups of the new structures were identified with the FINDSYM code~\cite{doi:10.1107/S0021889804031528}. Powder X-ray diffraction patterns were simulated by PowderCell~ \cite{Kraus:wi0185}. Using VASP,  constant volume molecular dynamics simulations were performed on {\it P}\,2$_1${\it /m}, {\it Pnma}, {\it Pm}, $\zeta$-{\it C}\,2{\it /m} and {\it P}\,6$_3$/{\it mmc} with PBE pseudopotential and Nos\'{e}-Hoover~\cite{10.1063/1.447334,PhysRevA.31.1695} thermostat  at $\sim$~116~GPa and $\sim$~140~GPa at 300~K. The simulation supercells contained 192 atoms for {\it Pm} with $\Gamma$-point sampling and 64 atoms for the other four structures with 2$\times$2$\times$2 BZ sampling grids. In addition, a supercell of 48 atoms for {\it Pm} with a 2$\times$2$\times$2 BZ grid was included.  A cutoff energy of 1000~eV and an integration time step of 0.75~fs were used.

Note that theoretical generation of the $\varepsilon$-O$_8$ structure above 50~GPa using the PAW method implemented in VASP is impossible, as the structure spontaneously converts to the metallic $\zeta$-{\it C}\,2{\it /m}~\cite{PhysRevB.76.064101}. In the present work, $\varepsilon$-O$_8$  between 30 and 140~GPa is constructed by restricting the electronic occupancies so that the lowest 24 electronic bands remain occupied, thus ensuring that $\varepsilon$-O$_8$  remains a semiconductor. We have confirmed that the HSE06-relaxed $\varepsilon$-O$_8$ remains  a semiconductor up to $\sim$~107~GPa at 0~K (see plots of the electron bands structure computed within PBE and HSE in Supplementary Material, Fig.~S1 (a) \& (b)). At finite temperature, band broadening due to the atomic motion should lead to gap closure at lower pressure, in better agreement with the experimental value of 96~GPa~\cite{PhysRevLett.74.4690} measured at room temperature. 

In addition to the previously suggested {\it C}\,2{\it /m} and {\it C}\,2{\it /c} candidate structures for the $\zeta$ phase~\cite{PhysRevB.76.064101}, our structure searches yield several new candidates, namely,  {\it R}\,-3{\it m}, {\it P}\,6$_3${\it /mmc}, {\it Cmcm}, {\it P}\,2$_1$, {\it Pnma}, {\it P}\,2$_1${\it /m}, and {\it Pm} (Fig.~\ref {All-St}). Their lattice parameters are reported in the Supplemental Material Table I. The first three structures are unlikely candidates for the $\zeta$ phase because their primitive cells have bases of only two and four atoms, which are insufficient to account for the measured Raman mode
mupltiplicities~\cite{PhysRevB.68.100102}. The remaining structures consist of 16, 8, 8, and 24 atoms, respectively. 

Figure~\ref{H-P_0K}(a) compares the PBE enthalpies of  the generated phases of oxygen with USPEX. As noted by Ma {\it et al.}~\cite{PhysRevB.76.064101}, {\it C}\,2{\it /m} is the dominant structure at the pressures considered here. However, the enthalpy differences between {\it Pnma}, {\it P}\,2$_1${\it /m}, {\it Pm}, and {\it C}\,2{\it /c} are less than 1~meV/atom. The discontinuity in the enthalpy curve of {\it Cmcm} is due to a transition to another structure with the same space group, {\it Cmcm}, above 90~GPa, as seen in the discontinuity of its lattice parameters at this pressure (see Fig. S2). All these structures are molecular with O$_2$ bond distance of about 1.18~{\AA} and the differences among them are in the relative arrangement of the molecules (see Fig.'s S3 to S8 for neighbor distance distribution analysis).  
Furthermore, all  candidate structures for the $\zeta$ phase are metallic, which is consistent with the experimental observation~\cite{PhysRevLett.74.4690}. Their electronic properties were evaluated at 0~K (see electronic DOS at 140~GPa in Fig.~S9) and for the {\it P}\,2$_{1}${\it /m}, {\it Pnma}, {\it Pm} and $\zeta$-{\it C}\,2{\it /m}  structures at 300~K  and $\sim$116~GPa. The latter was done by calculating average DOS over molecular dynamics simulations (see below) and employing the HSE06 functional at the electronic level. In all cases, the electronic band gap is closed.

After inclusion of the HSE06 hybrid functional in the geometrical optimization, the relative stability of the candidate structures changes significantly. The modified
enthalpy curves are shown in Fig.~\ref{H-P_0K}(b). In contrast to the PBE result, $\varepsilon$-O$_8$ is the energetically preferred structure below 111~GPa, while
{\it Pnma} and {\it Pm} are the lowest-enthalpy structures above this pressure. Within HSE06, the enthalpy of $\zeta$-{\it C}\,2{\it /m} becomes about 9~meV/atom higher than that of both  {\it Pnma} or {\it Pm} at 116~GPa. These are small enthalpy differences, which can be overcome by the phonon free energies even at room temperature.

\begin{figure} [t!]
 \centering 
\includegraphics[width=0.5\textwidth, clip]{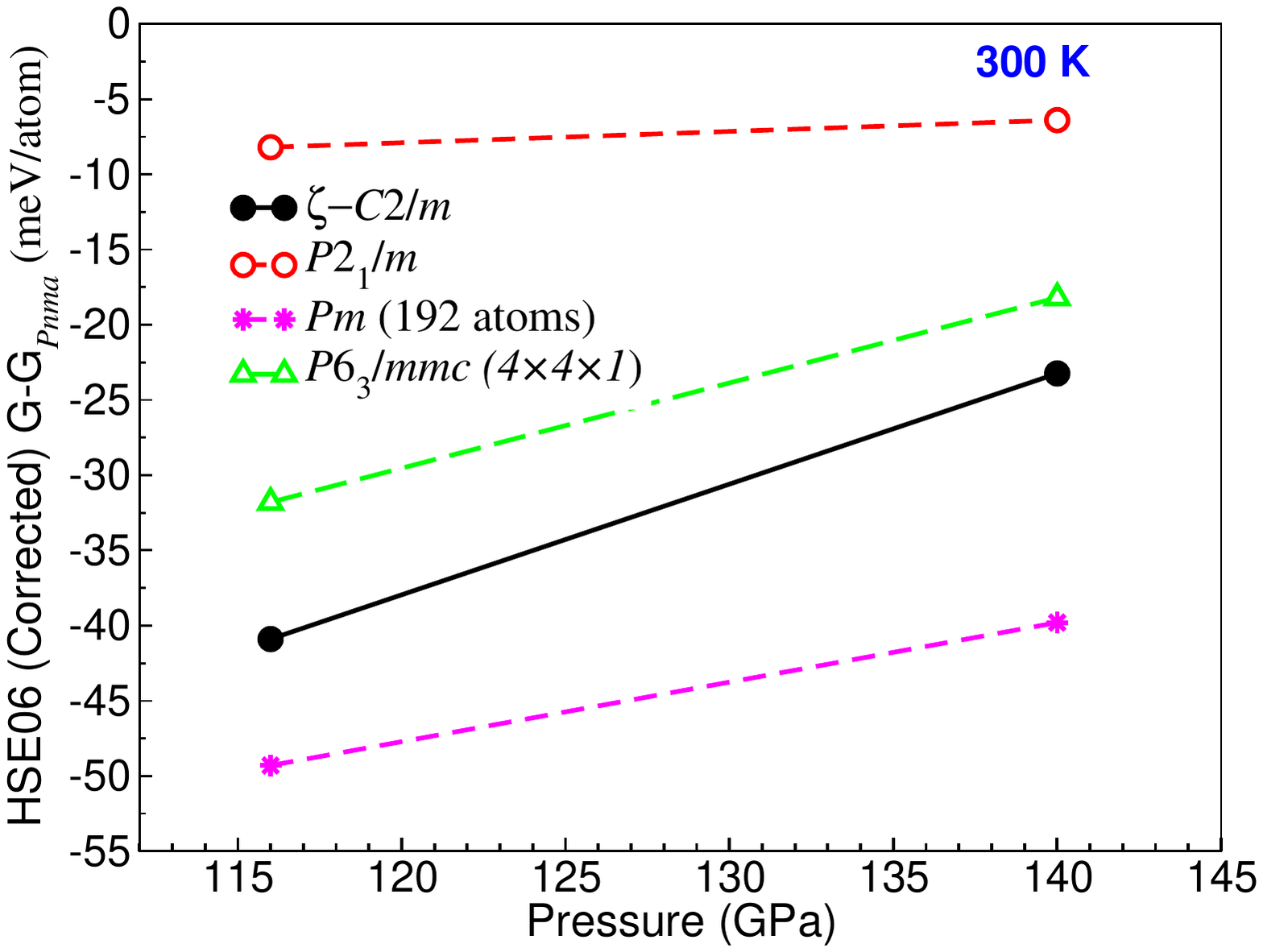}
\caption{HSE06-corrected Gibbs free energy difference relative to the {\it Pnma} structure at 300~K. \label{G-P_300K}} 
\end{figure}

To determine thermodynamic stability at finite temperature, we have performed constant-volume first principles molecular dynamics (FPMD) simulations at densities corresponding to pressures of about  116 and 140~GPa and temperature of 300~K for {\it P}\,2$_{1}${\it /m}, {\it Pnma}, {\it Pm}, {\it P}\,6$_3$/{\it mmc}  and $\zeta$-{\it C}\,2{\it /m}.  Equilibrium was established within the first 2~ps of the simulations and data was collected in the following 5~ps. The enthalpies of all structures, $H= U + PV$, were obtained directly from the FPMD at the level of DFT-PBE, with $U$ and $P$ as statistical averages of the instantaneous energy and pressure.   HSE06 corrections to the enthalpy were then made as $\Delta H_{HSE} = \langle E_{HSE} - E_{PBE}\rangle_{PBE}$, where  $E_{HSE}$ and $E_{PBE}$ are HSE and PBE energies, respectively, of atomic configurations taken from the FPMD trajectories, and  $\langle \cdots \rangle_{PBE}$ indicates statistical average within the PBE-FPMD ensemble. To obtain the Gibbs free energy, $G = H - TS$, the entropies, $S$,  of all structures were computed by integrating their vibrational density of states assuming a harmonic partition function. The vibrational density of states were obtained by Fourier transforming velocity autocorrelation functions from the FPMD trajectories. The relative HSE-corrected Gibbs free energies of the relevant structures are shown in Fig.~\ref{G-P_300K}.

\begin{figure} [t!]
 \centering 
\includegraphics[width=0.47\textwidth, clip]{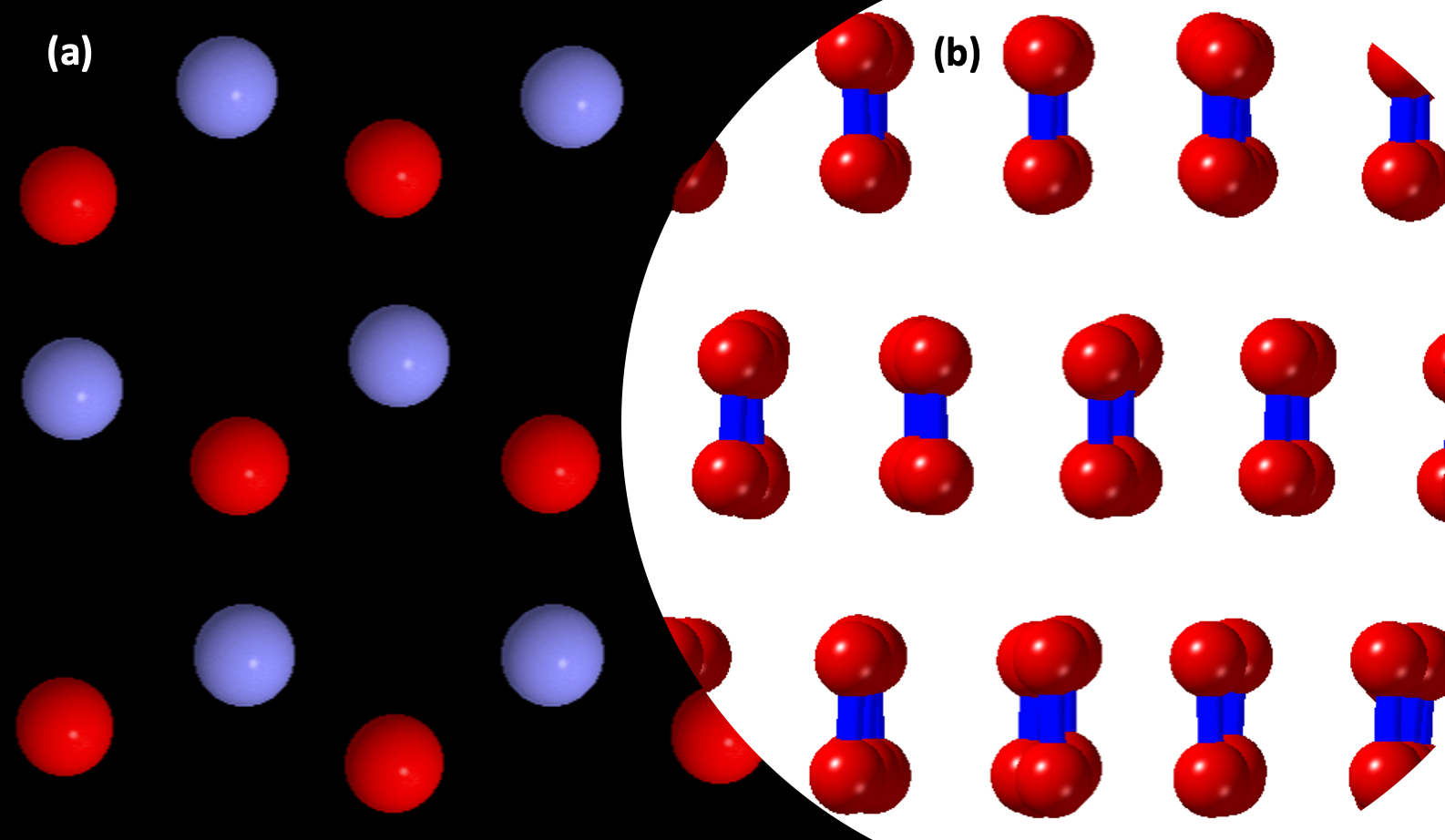}
\caption{View of the {\it Pm} structure at 116~GPa and 300~K (a) a long the $O_2$ bond and (b) side view. \label{Structure}}
\end{figure}

Note that the differences among the Gibbs free energies of the candidate structures are more significant than the 0~K enthalpy differences, which allows us to draw stronger conclusions for their stability. Based on the HSE06-corrected Gibbs free energy data presented in Fig.~\ref{G-P_300K}, we conclude that the structure for the $\zeta$ phase for oxygen is {\it Pm}.
The {\it Pm} structure has a near hexagonal closed packed  ({\it P}\,6$_3$/{\it mmc}) symmetry with a 24-atom unit cell with oxygen atoms at the 2c Wyckoff positions. Fig.~\ref{Structure} shows two neighboring basal planes of  {\it Pm} at $\sim$~116~GPa. Here a layer is shown in blue and the next one in red; the layers have $ABAB$ stacking.  

 In what follows, we compare properties of the energetically most competitive structures to experimental measurements of $\zeta$-oxygen phase. 

\begin{figure}[t!]
\includegraphics[scale=0.32,trim = 1mm 10mm 1mm 25mm, clip]{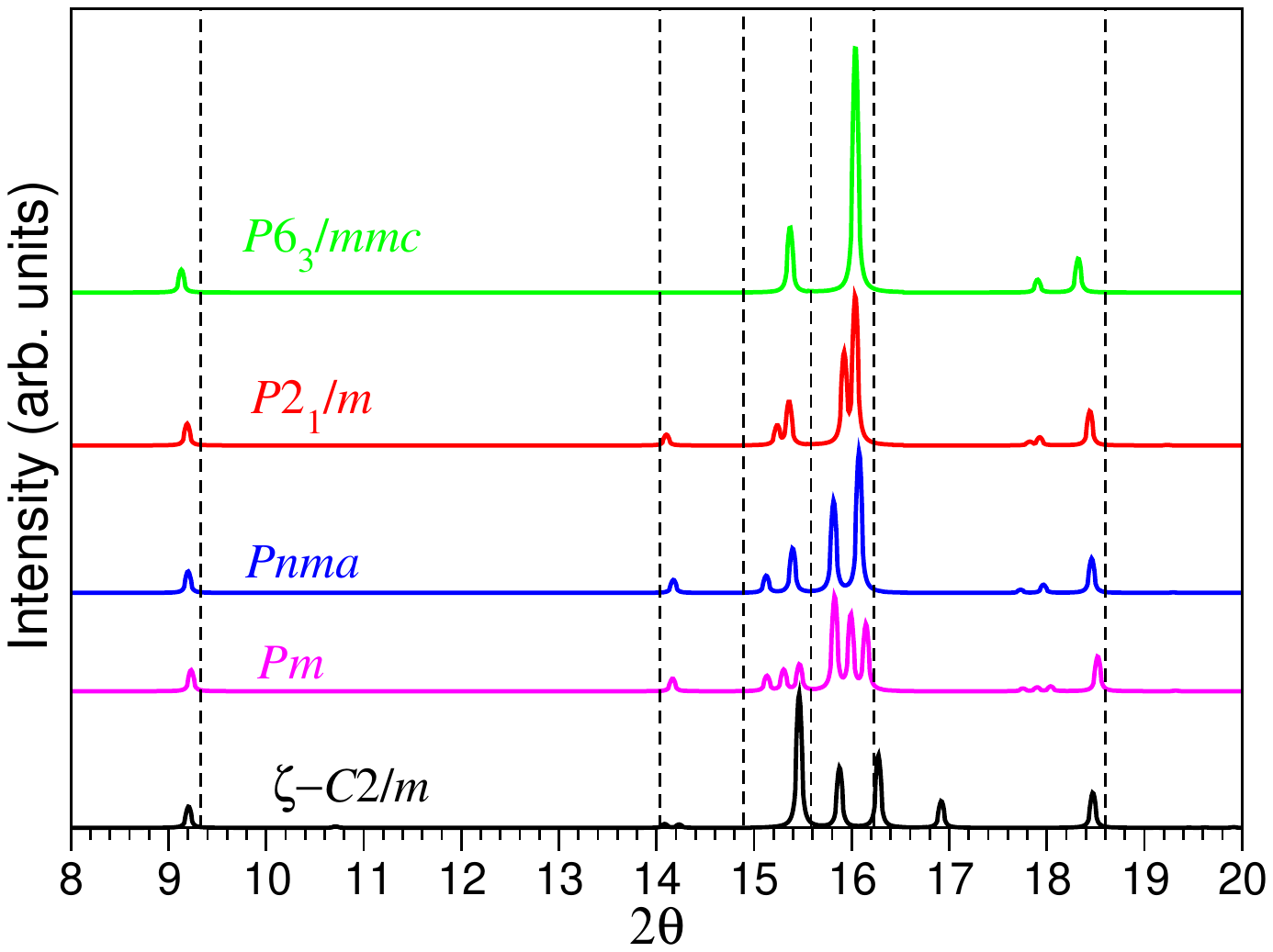}
\caption{Simulated powder  X-ray diffraction for HSE06-relaxed structures at 116~GPa. Experimental peaks are shown with dashed black lines \cite{PhysRevLett.74.4690}. \label{XRD}}
\end{figure}

 Simulated powder X-ray diffraction patterns for the HSE06-relaxed {\it P}\,2$_{1}${\it /m}, {\it Pnma}, {\it Pm}, $\zeta$-{\it C}\,2{\it /m} and {\it P}\,6$_3$/{\it mmc} at 116~GPa are shown in in Fig.~\ref{XRD}, along with the  X-ray peak positions  (dashed black lines) from experimental measurements obtained at the same pressure~\cite{PhysRevLett.74.4690}. The patterns for {\it P}\,2$_{1}${\it /m} and {\it Pnma} are almost identical to each other, as these two structures are nearly isostructural.  Previous analysis of X-ray data~\cite{PhysRevLett.74.4690, PhysRevLett.102.255503} has suggest that the structure of the $\zeta$ phase is monoclinic {\it C}\,2{\it /m}. Additionally, it has been suggested that the experimental peak at 14.9$^\circ$ belongs to the $\varepsilon$ phase, as the $\zeta$ and  $\varepsilon$ coexist from 96 to 128~GPa. However, we see here that all five structures show a reasonable agreement with the experimental data, though none of them matches the data perfectly. The measured~\cite{PhysRevLett.74.4690} X-ray peaks between 14$^\circ$ and 16$^\circ$ are rather broad and overlapping. One can argue that there are multiple peaks in this region, which are difficult to separate, in which case {\it Pm} provides the best agreement with the data. 

Metallic oxygen in the $\zeta$ phase has been found to be superconducting with a measured~\cite{Shimizu:1998} critical temperature ($T_c$) of 0.6~K. Here we have estimated the values
of $T_c$ for the HSE06-relaxed {\it P}\,2$_{1}${\it /m}, {\it Pnma}, {\it Pm}, $\zeta$-{\it C}\,2{\it /m} and {\it P}\,6$_3$/{\it mmc} structures at 116~GPa. They were calculated using
the ABINIT~\cite{Gonze20092582} package, with the McMillan's formula~\cite{PhysRevB.12.905} and  an average electron-electron Coulomb interaction length of  0.136. The $T_c$
estimates for the above phases are 30, 1.6, 6.2, 0.02, and 9.4~K, respectively. The computed $T_c$ for {\it Pnma} (1.6~K) and $\zeta$-{\it C}\,2{\it /m} (0.02~K) are closest to the experimental value. However, except for  {\it P}\,2$_{1}${\it /m}, the differences are small and the McMillan formula estimate is insufficient to discriminate among them, while more accurate superconducting calculations are beyond the scope of this paper.

Raman measurements on the $\zeta$ phase~\cite{PhysRevB.68.100102} indicate the presence of (at least) seven optical modes. {\it P}\,6$_3$/{\it mmc} has only three Raman active modes.  
{\it Pnma} and $\zeta$-{\it C}\,2{\it /m}  cannot account for all Raman modes either (see Fig. S10 in Supplementary Material). Finally,  {\it Pm} has 69 optical modes, which are both infrared and Raman active. It accounts for all measured Raman peaks, but exhibits extra modes as well. A possible explanation for this seeming discrepancy is that many of the optical modes are too weak to observe experimentally. Similarly to the x-ray data, the Raman peaks in the 400 to 900 cm$^{-1}$ range are broad and overlapping. This is the frequency region where most of the extra  {\it Pm} modes are.  Thus, another possibility is that individual modes are hard to identify from the available experimental data.

In summary, we have performed first-principles crystal structure searches identifying several new candidate structures for the $\zeta$ phase of oxygen. DFT calculations with a hybrid exchange functional (HSE06), which correctly predicts the $\varepsilon$ phase and the insulator-to-metal transition of oxygen at 0~K, significantly alter the relative stability of these structures compared to DFT-GGA. Even bigger effect has the inclusion of finite-temperature effects. The final energetic ordering is established from Gibbs free energies calculated using molecular dynamics simulations performed at room temperature and with HSE06 corrections at the level of thermodynamic perturbation theory. Based on this analysis, the equilibrium structure for the $\zeta$ phase of oxygen is a nearly hexagonal close-packed structure with the  {\it Pm} symmetry and a 24 atom-unit cell. Comparison with available experimental data shows a reasonable agreement with X-ray and Raman measurements -- as good, if not better, than that of the previously proposed {\it C}\,2{\it /m} structure. Nevertheless, the agreement is not perfect. Higher resolution measurements will be useful for validating our prediction. A possible difficulty in collecting high quality experimental data probably comes from the presence of a mixed-phase state, as reported in Raman studies~\cite{PhysRevB.68.100102} in the 96-124~GPa range. Reaching higher pressures could mitigate this problem.

\section*{Acknowledgments}
S.F.E. and V.A. contributed equally to this work. S.F.E. acknowledges support from King Fahd University of Petroleum and Minerals, Saudi Arabia and the Deanship of Scientific Research through Project No. SR211002. Computational resources were provided by ACEnet and the Canadian Foundation for Innovation. The work at LLNL was performed under the auspices of the U.S. DOE, Contract No. DE-AC52-07NA27344.\\
\textbf{Supplementary Information:\\} 
Supplementary Information accompanies this 
paper at https://journals.aps.org/prl/xxx

\bibliography{ref.bib}
\end{document}